# Thermal rectification in a bilayer wall: coupled radiation and conduction heat transfer

Hamou Sadat[1] and  Vital Le Dez

Institut P', Université de Poitiers, Centre National de la Recherche Scientifique, 2 Rue Pierre Brousse, Bâtiment B25, TSA 41105, 86073 Poitiers Cedex 9, France

**Abstract**: Thermal rectification can occur when heat is transferred asymmetrically. We consider in this work, the case of a composite wall with one conductive and radiative layer. Both Rosseland and surface radiation approximations are used for taking into account radiative transfer. Analytical expressions for the rectification factor are given. It is shown that this factor is bounded by a maximum value of around 2.62.

## I Introduction

Thermal rectification has been the subject of numerous studies because of the potential applications in heat management [1]. This phenomenon which has been first described in [2] can appear in heat radiation, heat convection and heat conduction as well. Radiative thermal rectifiers have been studied in near field and far field radiation. Rectification can be achieved in a device with two selective emitters one of which has radiative properties strongly dependent on temperature whereas the other one has temperature independent properties [3]. In order to enhance the rectification ratio, phase change radiative thermal rectifier based on the phase transition of insulator-metal transition materials around the operating temperature has also been considered [4]. The rectification is realized by the phase-change of vanadium dioxide (VO2), which behaves as metal at high temperature (> 340 K) and insulator at low temperature (< 340 K).

---

Convective thermal diode walls where the heat flow is more intense in one direction than in the other has been studied in [5-8] where rectification factors as high as 8 have been reported. The idea of heat conduction thermal rectifier in a composite wall consisting of two materials, each having temperature dependant conductivity was presented in [9-10]. Some experimental results have also been discussed [11-13]. More recently, it has been shown [14] that the maximum rectification factor in a composite wall consisting of two solid materials, each having thermal conductivities with different temperature linear dependences, is equal to 3. In this paper we focus on steady coupled heat conduction and radiation problem within a one dimensional wall composed of two different materials one of which is a radiative participating and conductive medium while the other is an opaque wall with a constant thermal conductivity. We shall assume that the conductivity of the first wall is also constant and that radiative transfer can be described by the Rosseland approximation and therefore has a power three increase of its radiative conductivity with temperature. This variation of the total conductivity of the first wall with temperature let us expect a rectification factor of the order of that obtained in pure heat conduction problems. In the following sections, the problem is solved analytically and an expression for the rectification factor is given. It is shown that the rectification factor has a maximum value of 2.6 which is indeed of the order of what is obtained in pure heat conduction problems. In order to examine the influence of the extinction coefficient, we have also considered the limiting case of an optically thin media where only surface radiation is taken into account in the first layer. Here again, it is found that rectification factor is bounded by the same maximum value.

**II Heat conduction model and radiation Rosseland approximation**

Let us consider a one-dimensional heat exchange problem with two solid different materials as depicted in figure 1. The first domain is a hot emitting and highly absorbing grey semi-transparent medium characterized by its absorption coefficient $\kappa$, while the other domain is made of an opaque conductive medium. The non-dimensional width of each of them is 0.5 giving a total thickness of 1. The thermal conductivities $\lambda_1$ and $\lambda_2$ of the two materials are both supposed to be constant. The first material is a participating medium with a high absorption coefficient so that radiative transfer within it can be described by the Rosseland approximation. Its total conductivity can therefore be written as: $\lambda(T) = \lambda_1 + \frac{16\sigma}{3\kappa}T^3$. Without loss of generality we will suppose that the boundary conditions are $T(0) = T_1$ and $T(1) = 0$. The problem with an imposed non zero temperature at $x = 1$ can be retrieved by a variable change. These boundary conditions which correspond to the forward case are inverted in the reverse case ($T(0) = 0$ and $T(1) = T_1$).

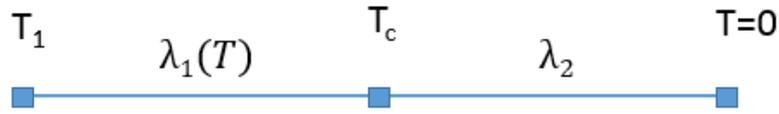

Figure 1: Heat conduction problem

The heat equation writes everywhere in the slab:

$$\frac{d}{dx}\left(\lambda \frac{dT}{dx}\right) = 0 \qquad (1)$$

The solution is simply given by $\lambda \frac{dT}{dx} = \beta$ where $\beta$ is a constant depending on the boundary conditions.

**II.1 Forward case**

In the second domain $x \in \left[\frac{1}{2}, 1\right]$ where $\lambda = \lambda_2$, the differential equation for the temperature field has the simple general solution $T = \frac{\beta}{\lambda_2}x + \gamma$, $\gamma$ being another constant to be determined, associated to the two boundary conditions $T(1) = 0$ and $T\left(\frac{1}{2}\right) = T_c$.

The determination of the two unknown constants easily gives the heat flux field in the domain: $q = 2\lambda_2 T_c$.

In the first region: $x \in \left[0, \frac{1}{2}\right]$, we can write:

$$\lambda \frac{dT}{dx} = \lambda_1 \frac{d}{dx}\left(T + \frac{4\sigma}{3\kappa\lambda_1} T^4\right) \quad (2)$$

The heat equation has therefore the general solution $T + \frac{4\sigma}{3\kappa\lambda_1} T^4 = \frac{\beta}{\lambda_1} x + \gamma$. The two boundary conditions $T(0) = T_1$ and $T\left(\frac{1}{2}\right) = T_c$ allow the determination of the two constants, which leads to the heat flux:

$$q = -2\lambda_1 \left[T_c \left(1 + \frac{4\sigma}{3\kappa\lambda_1} T_c^3\right) - T_1 \left(1 + \frac{4\sigma}{3\kappa\lambda_1} T_1^3\right)\right] \quad (3)$$

The continuity of the heat flux at $x = \frac{1}{2}$ gives the algebraic equation verified by the interface temperature:

$$T_c^4 + \frac{3\kappa\lambda_1}{4\sigma}\left(1 + \frac{\lambda_2}{\lambda_1}\right) T_c - \frac{3\kappa\lambda_1 T_1}{4\sigma}\left(1 + \frac{4\sigma}{3\kappa\lambda_1} T_1^3\right) = 0 \quad (4)$$

This 4$^{\text{th}}$ degree equation can exactly be solved by the Ferrari's method, associated to the resolvant third degree equation, and the temperature $T_c$ at the interface between the two media can be expressed as

$$T_c = \frac{1}{2}\left(\sqrt{\frac{2|q|}{\sqrt{t}} - t} - \sqrt{t}\right) \quad (5)$$

where $t = \sqrt[3]{\frac{q^2 + \sqrt{q^4 + 4\left(\frac{4r}{3}\right)^3}}{2}} - \sqrt[3]{\frac{\sqrt{q^4 + 4\left(\frac{4r}{3}\right)^3} - q^2}{2}} > 0$, the two parameters $q$ and $r$ being given by $q = -\frac{3\kappa\lambda_1}{4\sigma}\left(1 + \frac{\lambda_2}{\lambda_1}\right)$ and $r = \frac{3\kappa\lambda_1 T_1}{4\sigma}\left(1 + \frac{4\sigma}{3\kappa\lambda_1} T_1^3\right)$. Then the heat flux of the forward case is simply

$$q^+ = \lambda_2 \left( \sqrt{\frac{2|q|}{\sqrt{t}} - t} - \sqrt{t} \right) \quad (6)$$

**II.2 Reverse case**

In the reverse case, the general solution obtained in the forward case remains valid, with $T = \frac{\beta}{\lambda_2} x + \gamma$ for $x \in \left[\frac{1}{2}, 1\right]$ and $T + \frac{4\sigma}{3\kappa\lambda_1} T^4 = \frac{\beta'}{\lambda_1} x + \gamma'$ for $x \in \left[0, \frac{1}{2}\right]$.

Application of the boundary conditions sets leads to the two expressions of the heat conductive flux as $q = -2\lambda_2(T_1 - T_c)$ in the 2$^{nd}$ region and $q = -2\lambda_1 T_c \left(1 + \frac{4\sigma}{3\kappa\lambda_1} T_c^3\right)$ in the first one. Similarly to what precedes, the heat flux continuity gives the algebraic 4$^{th}$ degree equation verified by the interface temperature under the form:

$$T_c^4 + \frac{3\kappa\lambda_1}{4\sigma}\left(1 + \frac{\lambda_2}{\lambda_1}\right) T_c - \frac{3\kappa\lambda_2 T_1}{4\sigma} = 0 \quad (7)$$

Defining the parameter $s = \frac{3\kappa\lambda_2 T_1}{4\sigma}$ gives the solution $u$ of the resolvant third degree equation: $u = \sqrt[3]{\frac{q^2 + \sqrt{q^4 + 4\left(\frac{4s}{3}\right)^3}}{2}} - \sqrt[3]{\frac{\sqrt{q^4 + 4\left(\frac{4s}{3}\right)^3} - q^2}{2}} > 0$. The temperature $T_c$ at the interface between the two media and the heat flux of the reverse case are finally given by the following expressions:

$$T_c = \frac{1}{2}\left( \sqrt{\frac{2|q|}{\sqrt{u}} - u} - \sqrt{u} \right) \quad (9)$$

$$q^- = \lambda_2 \left( \sqrt{\frac{2|q|}{\sqrt{u}} - u} - \sqrt{u} - 2T_1 \right) \quad (10)$$

**II.3 Rectification factor**

The steady state heat fluxes in the forward and the reverse cases given by relations (6) and (10) allow us to define the rectification factor as the ratio $R = \left|\frac{q^+}{q^-}\right|$. This factor is given by the following formula:

$$R = \frac{\sqrt{\frac{2|q|}{\sqrt{t}} - t} - \sqrt{t}}{2T_1 + \sqrt{u} - \sqrt{\frac{2|q|}{\sqrt{u}} - u}} \quad (11)$$

Let us define a normalization temperature as $T_m^3 = \frac{3\kappa\lambda_1}{4\sigma}$, and the two parameters $\tau = \frac{T_1}{T_m}$ and $\rho = \frac{\lambda_2}{\lambda_1}$. With these notations, the coefficients of the 4$^{\text{th}}$ degree interface temperature equations for the forward and reverse cases are given by: $|q| = \alpha T_m^3$, $r = \beta T_m^4$ and $s = \gamma T_m^4$ where $\alpha = 1 + \rho$, $\beta = \tau(1 + \tau^3)$ and $\gamma = \rho\tau$. Introducing the two supplementary quantities:
$x = \sqrt[3]{\frac{\alpha^2 + \sqrt{\alpha^4 + 4\left(\frac{4\beta}{3}\right)^3}}{2}} - \sqrt[3]{\frac{\sqrt{\alpha^4 + 4\left(\frac{4\beta}{3}\right)^3} - \alpha^2}{2}}$ and $y = \sqrt[3]{\frac{\alpha^2 + \sqrt{\alpha^4 + 4\left(\frac{4\gamma}{3}\right)^3}}{2}} - \sqrt[3]{\frac{\sqrt{\alpha^4 + 4\left(\frac{4\gamma}{3}\right)^3} - \alpha^2}{2}}$, allows writing the rectification factor as a function of the two independent non dimensional parameters $\rho$ and $\tau$ under the form:

$$R = \frac{\sqrt{\frac{2\alpha}{\sqrt{x}} - x} - \sqrt{x}}{2\tau + \sqrt{y} - \sqrt{\frac{2\alpha}{\sqrt{y}} - y}} \quad (12)$$

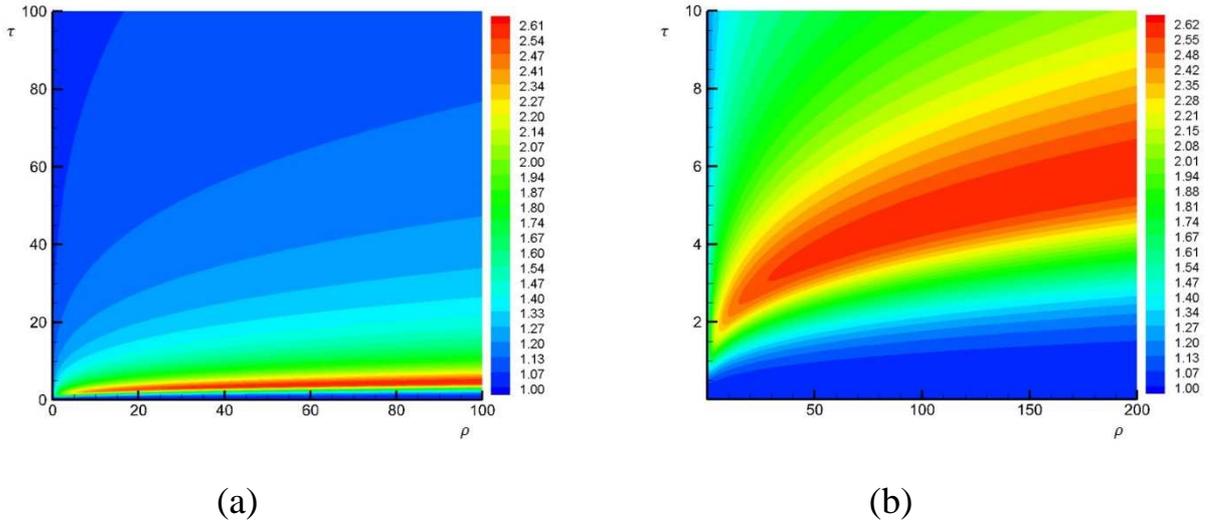

Figure 2: Rectification factor versus the two parameters $\rho$ and $\tau$

It is possible to show by taking the limit of the previous expression that $R \to 1$ for $\to 0$. We present in Fig. 2 the rectification factor as a function of $\rho$ and $\tau$. The first map on Fig. 2a shows the general representation of the rectification factor for a large spectrum of the two parameters $\rho$ and $\tau$, while the Fig. 2b focuses on a smaller region of small $\tau$ where the rectification factor takes significant values greater than 1.

As it can be observed, there exists a thin band of increasing expansion with $\rho$ of higher values of $R$ limited by to areas of decreasing values tending towards 1. The maximum value of the rectification factor is seen to be around: $R_m \approx 2.62$. The observation of the numerical results shows that if $\rho \geq 3$ then $R \geq 2$ if $0.73\rho^{\frac{1}{3}} \leq \tau \leq 1.47\rho^{\frac{1}{3}}$. In particular, it can be seen that $R \leq 1.5$ as soon as $\tau \geq 15$ for reasonable $\rho$ and that $R \to 1$ for $\tau \to +\infty$ for any $\rho$.

As a matter of example, one can see that a value of the rectification factor $R$ around 2 can be reached only for $\tau \geq 1.3$ which implies that the chosen temperature $T_1$ must be high enough if the absorption coefficient is not too small: if one chooses $T_1 = 1000K$, one obtains $T_m \approx 770K$ and $\kappa\lambda_1 \approx 34.5$. This case corresponds to a high absorption coefficient and a small thermal conductivity $\lambda_1$ with a large spectrum of conductivity of the second layer: $\lambda_2 \geq 2\lambda_1$.

**III Heat conduction model and thin media radiation approximation**

In the case of a thin radiation approximation with an absorption coefficient tending towards 0 (surface radiation), conservation of energy in medium 1 leads to write that the total heat flux $q_t = q_r + q_c$, sum of the radiative flux and the conductive one, is constant across the medium. This implies that:

$$\frac{\sigma \varepsilon_1 \varepsilon_c (T_1^4 - T_c^4)}{1 - \rho_1 \rho_c} - \lambda_1 \frac{dT}{dx} = \omega \quad (13)$$

where $\omega$ is a constant depending on the boundary conditions.

**III.1 Forward case**

In the first domain $x \in \left[0, \frac{1}{2}\right]$, equation (13) can be solved as follows:

$$T(x) = \frac{1}{\lambda_1} \left[ \frac{\sigma \varepsilon_1 \varepsilon_c (T_i^4 - T_c^4)}{1 - \rho_1 \rho_c} - \omega \right] x + \beta \quad (14)$$

where $\beta$ is a constant to be determined. The boundary conditions, $T\left(\frac{1}{2}\right) = T_c$ and $T(0) = T_1$, easily lead to: $\omega = \frac{\sigma \varepsilon_1 \varepsilon_c (T_1^4 - T_c^4)}{1 - \rho_1 \rho_c} - 2\lambda_1 (T_c - T_1)$, and the total flux in domain 1 writes:

$$q_t = \frac{\sigma \varepsilon_1 \varepsilon_c (T_1^4 - T_c^4)}{1 - \rho_1 \rho_c} - 2\lambda_1 (T_c - T_1) \quad (15)$$

There is no modification to the previous case in the second domain $x \in \left[\frac{1}{2}, 1\right]$ where the total flux consists only in a conductive one, given by $q = 2\lambda_2 T_c$. Continuity of the heat flux at the interface leads to the 4$^{th}$ degree equation verified by the temperature $T_c$:

$$T_c^4 + \frac{2\lambda_1(1-\rho_1\rho_c)}{\varepsilon_1\varepsilon_c\sigma}\left(1+\frac{\lambda_2}{\lambda_1}\right)T_c - \frac{2\lambda_1(1-\rho_1\rho_c)}{\varepsilon_1\varepsilon_c\sigma}\left[1+\frac{\varepsilon_1\varepsilon_c\sigma}{2\lambda_1(1-\rho_1\rho_c)}T_1^3\right] = 0 \qquad (16)$$

which is exactly similar to the equation obtained in the Rosseland approximation, with the substitution: $\kappa \leftrightarrow \frac{8(1-\rho_1\rho_c)}{3\varepsilon_1\varepsilon_c}$.

**III.2 Reverse case**

In the first domain $x \in \left[0, \frac{1}{2}\right]$, the total flux writes: $\omega = -\frac{\sigma\varepsilon_1\varepsilon_c T_c^4}{1-\rho_1\rho_c} - \lambda_1 \frac{dT}{dx}$, and the temperature variation is given by:

$$T(x) = -\frac{1}{\lambda_1}\left(\frac{\sigma\varepsilon_1\varepsilon_c T_c^4}{1-\rho_1\rho_c} + \omega\right)x + \beta \qquad (17)$$

The boundary conditions, $T\left(\frac{1}{2}\right) = T_c$ and $(0) = 0$, now lead to the total flux expression:

$$q_t = -2\lambda_1 T_c\left[1+\frac{\varepsilon_1\varepsilon_c\sigma}{2\lambda_1(1-\rho_1\rho_c)}T_c^3\right] \qquad (18)$$

Similarly there is no modification in the area $\frac{1}{2} \leq x \leq 1$ where the flux writes $q_t = -2\lambda_2(T_1 - T_c)$. The equation verified by the interface temperature $T_c$ is :

$$T_c^4 + \frac{2\lambda_1(1-\rho_1\rho_c)}{\varepsilon_1\varepsilon_c\sigma}\left(1+\frac{\lambda_2}{\lambda_1}\right)T_c - \frac{2\lambda_2(1-\rho_1\rho_c)T_1}{\varepsilon_1\varepsilon_c\sigma} = 0 \qquad (19)$$

The same substitution $\kappa \leftrightarrow \frac{8(1-\rho_1\rho_c)}{3\varepsilon_1\varepsilon_c}$ allows to retrieve the equation verified in the Rosseland approximation.

Then introducing the normalisation temperature $T_m^3 = \frac{2\lambda_1(1-\rho_1\rho_c)}{\varepsilon_1\varepsilon_c\sigma}$, so as the independant parameters $\tau = \frac{T_1}{T_m}$ and $\rho = \frac{\lambda_2}{\lambda_1}$, allows writing the rectification factor as:

$$R = \frac{\sqrt{\frac{2\alpha}{\sqrt{x}}-x}-\sqrt{x}}{2\tau+\sqrt{y}-\sqrt{\frac{2\alpha}{\sqrt{y}}-y}} \qquad (20)$$

where $\alpha$, $x$ and $y$ have exactly the same expressions as in the Rosseland approximation frame. One concludes that the map representing the rectification factor for the thin media approximation is identical to the one obtained for the Rosseland approximation, and that the maximal value of the rectification factor is $R_m \approx 2.62$.

In the case of black boundaries, the normalisation temperature is: $T_m^3 = \frac{2\lambda_1}{\sigma}$. For a thin medium with $\lambda_1 = 1\ W.m^{-1}.K^{-1}$, this leads to $T_m \approx 328K$ and $\tau \approx 3$ when $T_1 = 1000K$ and relatively high values of $R$ around 2 can be obtained for media with $\lambda_2 \geq 2$, the maximum $R_m$ being reached for $\lambda_2 \approx 30$. If $\lambda_1 = 0.1\ W.m^{-1}.K^{-1}$, $T_m \approx 152K$ and $\tau \approx 6.6$, leading to the maximal value of $R$ reached for media with $\lambda_2 \geq 18$.

**IV Conclusion**

In this study we developed an analytical expression for the rectification factor in a bilayer wall when radiation and conduction are coupled in the first layer. By using the Rosseland approximation and by considering surface radiation in the first layer where conduction is also taken into account while the second layer is only conductive, we have given an analytical expression for the rectification factor. A theoretical maximum value has been numerically found, which can hardly be obtained in the Rosseland approximation frame, except for a very small thermal conductivity in the hot domain. On the contrary for the thin media approximation, values close to the theoretical maximum rectification factor can be obtained in less restrictive conditions.